\documentclass[aps,prc,twocolumn, twoside, amsmath,superscriptaddress]{revtex4-1}

\usepackage[utf8]{inputenc}
\usepackage[T1]{fontenc}

\usepackage{graphicx}
\usepackage{rotating}
\usepackage{xcolor}
\usepackage{pifont}
\usepackage{textcomp}
\usepackage{ifthen}
\usepackage{multirow}

\usepackage{isotope}

\usepackage{scalefnt}

\usepackage{times}
\usepackage{txfonts}

\usepackage{amsmath}
\usepackage{amssymb}
\usepackage{amsfonts}
\usepackage{braket}
\usepackage{bm}

\definecolor{indianred}{rgb}{0.86, 0.08, 0.24}
\definecolor{royalblue}{rgb}{0.25, 0.41, 0.88}
\definecolor{darkorange}{rgb}{1.0, 0.55, 0}
\definecolor{mediumseagreen}{rgb}{0.24, 0.70, 0.44}
\definecolor{purple}{rgb}{0.5, 0, 0.5}
\definecolor{cyan3}{rgb}{0, 0.80, 0.80}

\definecolor{plot1}{rgb}{0.86, 0.08, 0.24}
\definecolor{plot2}{rgb}{0.25, 0.41, 0.88}
\definecolor{plot3}{rgb}{1.0, 0.55, 0}
\definecolor{plot4}{RGB}{61,153,86}

\newcommand{\la}{\langle}
\newcommand{\ra}{\rangle}

\newcommand{\symboltriangleup}[1][black]{{\color{#1}\scalefont{0.9}{\raisebox{1.5ex}{\begin{turn}{180}$\blacktriangledown$\end{turn}}}}}

\newcommand{\symbolcircle}[1][black]{{\color{#1}\scalefont{0.75}\ding{108}}}

\newcommand{\symboltriangleopen}[1][black]{{\color{#1}$\bm{\bigtriangleup}$}}

\newcommand{\symbolcircleopen}[1][black]{{\color{#1}\scalefont{0.75}\LARGE$\circ$}}

\newcommand{\symboldiamondsym}[1][black]{{\color{#1}\scalefont{0.75}\raisebox{-.2ex}{\begin{turn}{45}$\blacksquare$\end{turn}}}}

\newcommand{\symbolstar}[1][black]{{\color{#1}\raisebox{0.2ex}{$\bigstar$}}}

\newcommand{\redcircle}{{\scalefont{0.9}\symbolcircle[indianred]}}
\newcommand{\redcircleopen}{{\scalefont{0.9}\symbolcircleopen[indianred]}}
\newcommand{\reddiamond}{{\scalefont{0.9}\symboldiamondsym[indianred]}}

\newcommand{\bluetriangleupopen}{{\scalefont{0.9}\symboltriangleopen[royalblue]}}
\newcommand{\bluetriangleup}{{\scalefont{0.9}\symboltriangleup[royalblue]}}

\newcommand{\bluestar}{{\scalefont{0.9}{{\scalefont{0.8}\symbolstar[royalblue]}}}}

\newcommand{\orangetriangleup}{{\scalefont{0.9}{\scalefont{0.8}\symboltriangleup[darkorange]}}}

\newcommand{\orangestar}{{\scalefont{0.9}\scalefont{0.8}\symbolstar[darkorange]}}

\newcommand{\greencircle}{{\scalefont{0.9}{\symbolcircle[plot4]}}}

\newcommand{\elem}[2]{\ensuremath{{}^{#2}\text{#1}}}

\begin{document}

\title{Hartree-Fock Many-Body Perturbation Theory for Nuclear Ground-States}

\author{Alexander Tichai}
\email{alexander.tichai@physik.tu-darmstadt.de}
\affiliation{Institut f\"ur Kernphysik, Technische Universit\"at Darmstadt, 64289 Darmstadt, Germany}

\author{Joachim Langhammer}
\affiliation{Institut f\"ur Kernphysik, Technische Universit\"at Darmstadt, 64289 Darmstadt, Germany}

\author{Sven Binder}
\affiliation{Department of Physics and Astronomy, University of Tennessee, Knoxville, TN 37996, USA}
\affiliation{Physics Division, Oak Ridge National Laboratory, Oak Ridge, TN 37831, USA}

\author{Robert Roth}
\email{robert.roth@physik.tu-darmstadt.de}
\affiliation{Institut f\"ur Kernphysik, Technische Universit\"at Darmstadt, 64289 Darmstadt, Germany}

\begin{abstract}
We investigate the order-by-order convergence behavior of many-body perturbation theory (MBPT) as a simple and efficient tool to approximate the ground-state energy of closed-shell nuclei. To address the convergence properties directly, we explore perturbative corrections up to 30$^{\text{th}}$ order and highlight the role of the partitioning for convergence. The use of a simple Hartree-Fock solution to construct the unperturbed basis leads to a convergent MBPT series for soft interactions, in contrast to, e.g., a harmonic oscillator basis. For larger model spaces and heavier nuclei, where a direct high-order MBPT calculation in not feasible, we perform third-order calculation and compare to advanced \emph{ab initio} coupled-cluster calculations for the same interactions and model spaces. We demonstrate that third-order MBPT provides ground-state energies for nuclei up into tin isotopic chain that are in excellent agreement with the best available coupled-cluster results at a fraction of the computational cost.

\end{abstract}


\maketitle

%%%%%%%%%%%%%%%%%%%%%%%%%%%%%%%%%%%%%%%%%%%%%%%%%%%%%%%%%%%%%%%
\paragraph*{Introduction.}
The solution of the Schr{\"o}dinger equation for atomic nuclei using realistic nuclear interactions is at the heart of \emph{ab initio} nuclear structure theory. In practice this problem is solved by constructing approximate methods for a truncated, i.e., finite-dimensional Hilbert space. However, for the calculation of ground-state energies of heavy nuclei significant algorithmic and computational efforts are needed. There exist a plethora of different \emph{ab initio} methods, e.g., coupled cluster (CC) theory \cite{DeHj04,BaRo07,HaPa10,KoDe04,PiGo09,BiLa14}, in-medium similarity renormalization group (IM-SRG) ~\cite{HeBo13,TsBo11,Mo15,He14,H15}, or self-consistent Green's function methods \cite{CiBa13,SoCi14,SoBa13}. However, it is desirable to have an alternative, light-weight framework available. A conceptually simple method to solve for the eigenenergies of a physical system is many-body perturbation theory (MBPT) \cite{Schr26,ShBa09,SzOs82}. A perturbative treatment is the standard approach for many problems from different fields of theoretical physics. The advantage of MBPT compared to other \emph{ab initio} approaches is its simplicity, which also allows for straightforward generalizations to excited states and open-shell nuclei~\cite{LaRo12} without the need of sophisticated equation-of-motion techniques. The reason why MBPT usually is not considered as \emph{ab initio} technique are convergence issues of the underlying perturbation series. Several studies of high-order MBPT based on Slater determinants constructed from harmonic oscillator (HO) single-particle states (HO-MBPT) have shown that the perturbation series is divergent in almost every case~\cite{LaRo12,RoLa10}. In such cases one heavily relies on the use of resummation techniques, e.g., Pad\'e approximants, that enable a robust extraction of observables although the perturbative expansion diverges~\cite{Bake65,BaGr96,RoLa10}. 

In this Letter, we formulate MBPT based on Hartree-Fock (HF) single-particle states (HF-MBPT), and, for the first time, investigate the convergence behavior of the perturbation series up to $30^{\text{th}}$ order.
We compare the ground-state energies of \elem{He}{4} and \elem{O}{16} to results from exact diagonalizations in the configuration interaction (CI) approach using the same model space \cite{RoLa11,BarNa13,MaAk13}. Based on the rapidly converging perturbation series resulting from the use of HF basis states, we study ground-state energies of selected closed-shell medium-mass and heavy nuclei at third-order MBPT, and compare to recent coupled cluster (CC) calculations~\cite{BiLa14}.

\paragraph*{The Nuclear Hamiltonian.}

For all following investigations we start from the chiral nucleon-nucleon (NN) interaction at next-to-next-to-next-to leading order (N$^3$LO) by Entem and Machleidt~\cite{EnMa03} combined with the three-nucleon (3N) interaction at next-to-next-to leading order N$^2$LO in its local form ~\cite{Na07} with three-body cutoff $\Lambda_{\text{3N}}=400 \,$MeV$/c$. Additionally, we use the similarity renormalization group (SRG) to soften the Hamiltonian through a continuous unitary transformation controlled by a flow parameter~\cite{BoFu07,HeRo07,RoRe08,RoCa13,JuMa13}. In principle this transformation induces beyond-3N operators up the mass number of the considered nucleus, which, however, we have to neglect.
To avoid the complication of dealing with explicit 3N interactions, we make use of the normal-ordered two-body approximation (NO2B) of the 3N interaction that has been found to be very accurate for medium-mass nuclei, see Refs.~\cite{RoBi12, BiLa13}. For the matrix-element preparation we adopt the procedure we introduced in Ref.~\cite{BiLa14}, i.e., in particular, we use the large SRG model-space and exploit the iterative scheme where necessary. Thus, the matrix elements and the treatment of the chiral NN+3N interaction are identical to Ref.~\cite{BiLa14} and we can compared directly to the CC results presented there.

%%%%%%%%%%%%%%%%%%%%%%%%%%%%%%%%%%%%%%%%%%%%%%%%%%%%%%%%%%%%%%%
\paragraph*{Many-Body Perturbation Theory.}\label{MBPT}

The essence of Rayleigh-Schr\"odinger perturbation theory is the definition of an additive splitting, referred to as partitioning, of a given Hamiltonian $H$ into an unperturbed part $H_0$ and a perturbation $W$. Introducing an auxiliary parameter $\lambda$ yields a one-parameter family of operators,
\begin{flalign}
%\label{eq:partitioning}
  H_{\lambda} = H_0 + \lambda W,
\end{flalign}
where the perturbation is defined by $W= H - H_0$. As ansatz for the solution of the eigenvalue problem of $H$ we take a power series expansion in terms of an auxiliary parameter $\lambda$, where the expansion coefficients are given by the energy corrections and state corrections, respectively. 
We choose $H_0$ to be the HF Hamiltonian arising from an initial NN+3N interaction. 
We have shown in Refs.~\cite{RoLa10, LaRo12} that high-order MBPT corrections are accessible by means of a recursive scheme, allowing for detailed investigations of the convergence characteristics of the perturbation series. In general we cannot expect that a perturbation series is convergent  \cite{BeWu69,BeWu73,BeOr99}, but one can exploit resummation-theory techniques to extract information on the observables of interest. There are different schemes and transformations that can be used to extract, e.g., the ground-state energy from a divergent expansion \cite{We89,Bo99,PeYo08}. Pad\'e approximants have proven to be particularly useful in the treatment of high-order HO-MBPT \cite{RoLa10,LaRo12}. Additionally, they are well-known to mathematicians especially in the field of convergence acceleration \cite{BaGr96,BeOr99,We89}. However, the calculation of energy corrections up to sufficiently high orders is only feasible for light nuclei due to increasing computational requirements. 
When proceeding to the medium-mass region one must choose a different strategy. Depending on the rate of convergence, one might expect low-order partial sums of the perturbation series to be a reasonable approximation to the exact ground-state energy. 
Having only low-order information available, resummation methods are less effective, because one can only construct a small number of approximants that yield valid approximations only if the transformed sequence converges sufficiently fast~\cite{LaRo12}. However, one alternative is to exploit the freedom in the partitioning, i.e., the choice of the unperturbed basis, to improve the convergence of the perturbation series.
For low-order calculations the corrections can be expressed in terms of the particle-hole formalism. Note that the Hartree-Fock energy  corresponds to the the first-order partial sum,
\begin{align}
 E_{\text{HF}} = E^{(0)} + E^{(1)} \;.
\end{align}
Therefore, the first contribution to the correlation energy appears in second-order HF-MBPT.
The second- and third-order partial sum for the ground-state energy with respect to HF basis for a two-body operator are given by \cite{RoPa06}
\begin{flalign}
\label{eq:thirdorder}
 \hspace{-0.5cm}E^{(2)}&=  \frac{1}{4} \sum_{ab}^{<\epsilon_{\text{F}}} \sum_{ij}^{>\epsilon_{\text{F}}}  \frac{\la ab| W |ij\ra\la ij|  W|ab\ra}{(\epsilon_a + \epsilon_b -\epsilon_i -\epsilon_j )} \notag \\
\hspace{-0.5cm}E^{(3)} &=  \frac{1}{8}  \sum_{abcd}^{<\epsilon_{\text{F}}} \sum_{ij}^{>\epsilon_{\text{F}}} \frac{\la ab |W|ij  \ra \la ij |W| cd \ra \la cd |W|ab \ra}{(\epsilon_a + \epsilon_b -\epsilon_c -\epsilon_d )(\epsilon_a + \epsilon_b -\epsilon_i -\epsilon_j )} \notag  \displaybreak[0] \\
 \hspace{-0.5cm}	      &+ \frac{1}{8} \sum_{ab}^{<\epsilon_{\text{F}}} \sum_{ijkl}^{>\epsilon_{\text{F}}} \frac{\la ab |W|ij \ra \la ij |W|kl \ra \la kl |W|ab \ra}{(\epsilon_a + \epsilon_b -\epsilon_i -\epsilon_j )(\epsilon_a + \epsilon_b -\epsilon_k -\epsilon_l )}  \notag\displaybreak[0] \\
 \hspace{-0.5cm}	      &+ \sum_{abc}^{<\epsilon_{\text{F}}} \sum_{ijk}^{>\epsilon_{\text{F}}}  \frac{\la ab |W|ij \ra \la cj |W|kb \ra \la ik |W|ac \ra}{(\epsilon_a + \epsilon_b -\epsilon_i -\epsilon_j )(\epsilon_a + \epsilon_c -\epsilon_i -\epsilon_k )}  .
\end{flalign}
%%%%%%%%
In the third-order energy correction the first, second and third term are called particle-particle (pp), hole-hole (hh) and particle-hole (ph) correction, respectively.
The $\epsilon_i$ correspond to the HF single-particle energies and all matrix elements are taken to be antisymmetrized. Summation indices $a,b,c,..$ correspond to particle indices, i.e., above the Fermi level $ \epsilon_{\text{F}}$, whereas $i,j,k,...$ correspond to hole indices ranging from lowest occupied single-particle states up to the Fermi level. The zero- and one-body parts of the normal-ordered Hamiltonian only enter in the first-order energy correction. Brillouin's theorem states that there is no mixing of the Hartree-Fock ground state with singly-excited determinants \cite{SzOs82} and by orthogonality the zero-body part is only present in the expectation value of the perturbation. 
In principle, the derivation of energy corrections beyond third order is straight forward. However, considering a diagrammatic approach in terms of Hugenholtz diagrams, the number of contributing diagrams at a given perturbation order $p$  increases rapidly \cite{OEIS} such that it becomes challenging to go beyond third-order in practice. Additionally, terms from higher-order corrections involve expressions that are notoriously hard to compute, because their effective implementation, e.g., by means of BLAS-enabled matrix operations, is not obvious. 
The computational power needed to perform third-order MBPT calculations up to the heavy-mass region can in principle be provided by a single computing node within $1 - 3\%$ of the computing time needed for state-of-the-art CC calculations.

\begin{figure}
\includegraphics[width=0.49\textwidth]{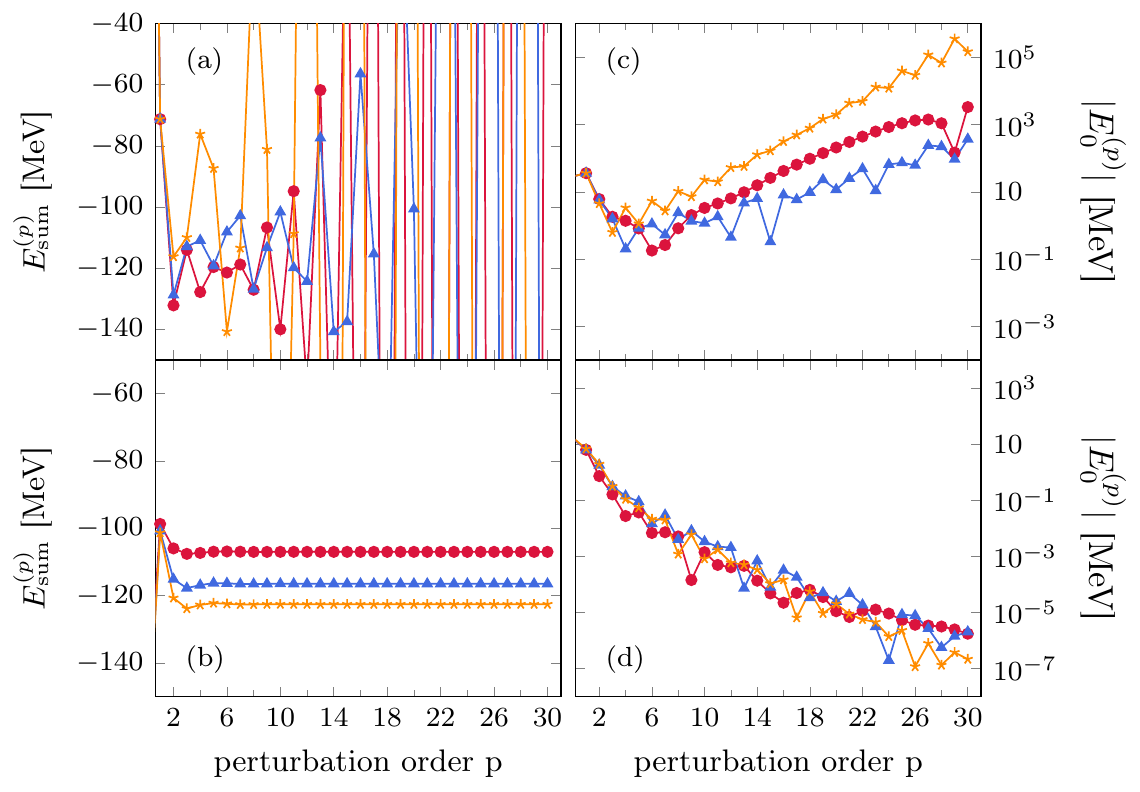}
\caption{Partial sums of $^{16}\text{O}$ in HO basis (a) and HF basis (b) for the NN+3N-full interaction with $\alpha=0.08\text{fm}^4$ and truncation parameters $N_{\text{max}}=2$ (\redcircle),\, $4$ (\bluetriangleup)  and $6$ (\orangestar). The corresponding energy corrections for each order are displayed in (c) and (d), respectively.  All calculations are performed at oscillator frequency $\hbar \Omega =24\,\text{MeV}$.} 
\label{convergence}
\end{figure}

%%%%%%%%%%%%%%%%%%%%%%%%%%%%%%%%%%%%%%%%%%%%%%%%%%%%%%%%%%%%%%%
\paragraph*{Convergence Characteristics of Hartree-Fock Many-Body Perturbation Theory.}\label{highorder}
We start with comparing perturbation series from HO- and HF-MBPT, and focus on their convergence characteristics and sensitivity to the SRG flow parameter.  In Fig.~\ref{convergence} we present a direct comparison of the order-by-order behavior for the two partitionings up to 30th order for \elem{O}{16}. For these high-order calculations we use an $N_{\text{max}}$-truncation of the many-body model space similar the no-core shell model (NCSM) \cite{BarNa13}. 
The left column of Fig.~\ref{convergence} shows the high-order partial sums and the right column the individual energy corrections for each order. 
Panel (a) shows the partial sums from HO-MBPT for a sequence of model spaces with fixed SRG flow parameter $\alpha=0.08\,\text{fm}^4$. The partial sums are divergent for every model space. The divergence is also apparent from panel (c) which reveals the exponentially increasing energy corrections. In contrast, panel (b) shows the partial sums arising from HF-MBPT that are convergent for all model spaces. Furthermore, the converged values agree with direct CI results. As seen in panel (d), the energy corrections are exponentially suppressed for higher orders, giving rise to a robust convergence. 

\begin{figure}
\includegraphics[width=0.49\textwidth]{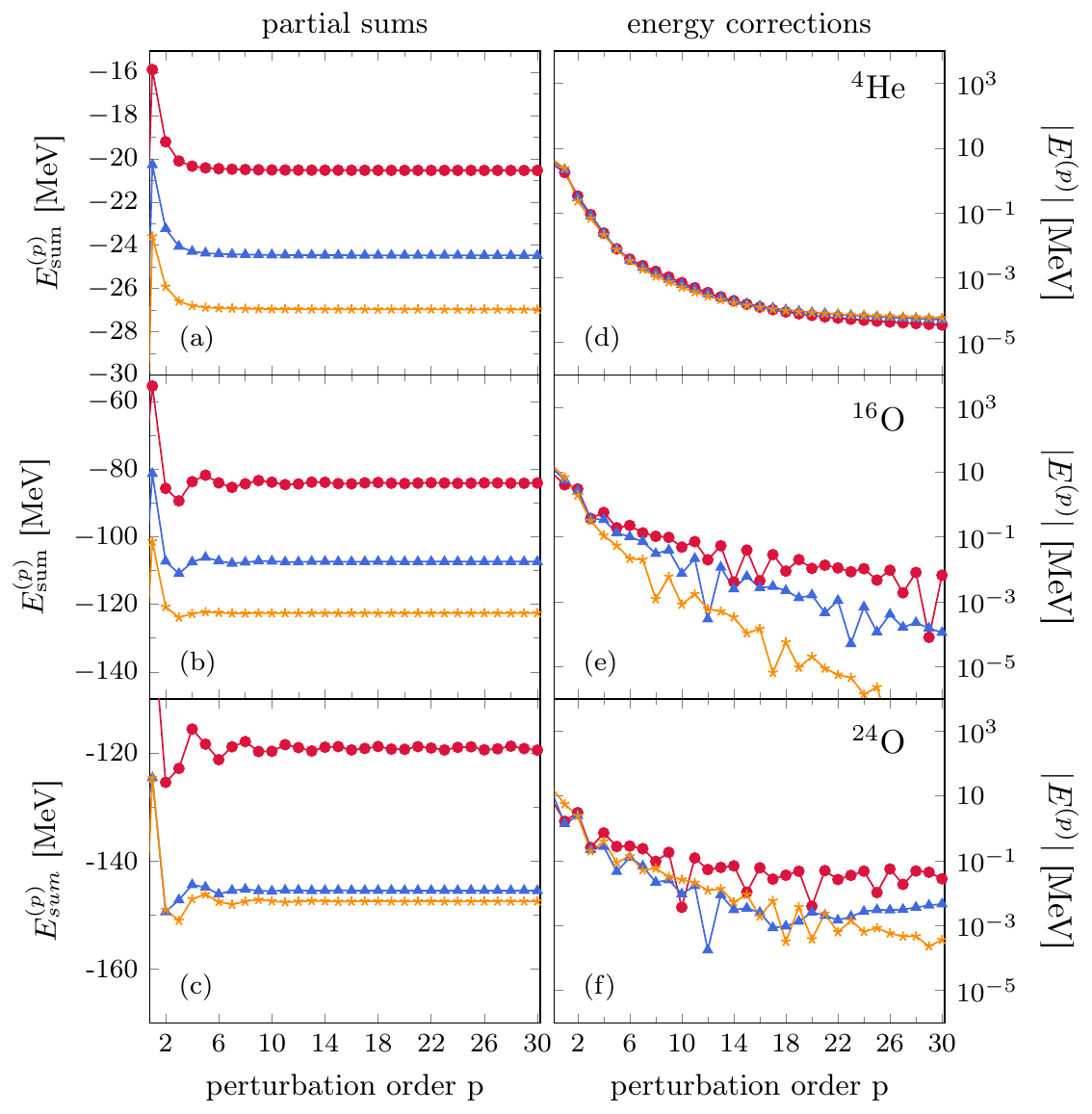}
\caption{Partial sums for varying flow parameters in HF-MBPT for \elem{He}{4} (a), \elem{O}{16} (b) and \elem{O}{24} (c). The corresponding energy corrections are given in (d), (e) and (f), respectively. The model space for the first and second panel are truncated at  $N_{\text{max}}=6$. The truncation for the third panel is given by $N_{\text{max}}=4$. The flow parameters for the different data sets are $\alpha=0.02 \,\text{fm}^4$ (\redcircle), $0.04 \,\text{fm}^4$ (\bluetriangleup) and $0.08 \,\text{fm}^4$(\orangestar).  All calculations use a NN+3N-full interaction with oscillator frequency $\hbar \Omega =24\,\text{MeV}$.}
\label{srgdependence}
\end{figure}

In Fig.~\ref{srgdependence} we show the high-order partial sums and energy corrections in HF-MBPT for different SRG flow parameters. Panels (a), (b) and (c) show the convergent perturbation series for $\elem{He}{4}$,$\elem{O}{16}$ and $\elem{O}{24}$, respectively. The calculations are performed for fixed $N_{\max}=6$ for $\elem{He}{4}$, $\elem{O}{16}$ and $N_{\max}=4$ for $\elem{O}{24}$ and the flow-parameter dependence of the absolute energies results from the varying degree of convergence with respect to the many-body model space. 

A more interesting flow-parameter dependence can be observed for the individual energy corrections in panels (d), (e), and (f). There is a clear dependence of the convergence rate on the flow parameter for the oxygen isotopes. For \elem{O}{16} the series converges exponentially in all three cases and the larger the flow parameter, i.e. the softer the Hamiltonian, the more rapid the convergence---as might be naively expected. For \elem{O}{24} the behavior is slightly more complicated. For the softest interaction with $\alpha=0.08\text{fm}^4$ there is still a clear exponential convergence. However, for the harder interactions, i.e., $\alpha=0.02, 0.04\text{fm}^4$, we observe no systematic decrease of the high-order perturbative contributions anymore, they remain approximately constant and cause a small-amplitude oscillatory behavior of the partial sums. However, even in these cases we can easily extract a robust estimate for the asymptotic value. In the case of $\elem{He}{4}$ the suppression is independent of $\alpha$ and we observe the same rapid convergence for all interactions. 
  
%%%%%%%%%%%%%%%%%%%%%%%%%%%%%%%%%%%%%%%%%%%%%%%%%%%%%%%%%%%%%%%
%
%%%%%%%%%%%%%%%%%%%%%%%%%%%%%%%%%%%%%%%%%%%%%%%%%%%%%%%%%%%%%%%
%
%
%%%%%%%%%%%%%%%%%%%%%%%%%%%%%%%%%%%%%%%%%%%%%%%%%%%%%%%%%%%%%%%

The numerical values of the partial sums for selected orders of HF-MPBT for the three nuclei and the different flow parameters are summarized in Tab.~\ref{hetable} together with the results of direct CI calculations for the same Hamiltonians and model spaces. The higher-order partial sums are in good agreement with the CI results---in most cases the deviation of the ground-state energy is much smaller than $0.1$\%. 

Based on  our detailed analysis of high-order HF-MBPT and due to the exponential suppression of the energy corrections, we can take low-order partial sums as a reasonable approximation to the converged results. This motivates the investigation of third-order partial sums for selected medium-mass and heavy closed-shell nuclei in the following. 

%%%%%%%%%%%%%%%%%%%%%%%%%%%%%%%%%%%%%%%%%%%%%%%%%%%%%%%%%%%%%%%
\begin{table}
\caption{Ground-state energies for \elem{He}{4}, \elem{O}{16} and \elem{O}{24} in units of [MeV] obtained in HF-MBPT for different orders up to $p=30$ and in CI calculations with NN+3N-full interactions for different flow paramters $\alpha$. The model spaces are truncated by $N_{\text{max}}=6$ for \elem{He}{4} and \elem{O}{16} and $N_{\text{max}}=4$ in the case of \elem{O}{24}. The HO frequency is $\hbar \Omega=24\,\text{MeV}$.}
\label{hetable}
\setlength{\tabcolsep}{7pt}
\begin{ruledtabular}
\begin{tabular}{c  c  | c | c | c }
& &  \multicolumn{3}{c}{$\alpha\;\;  [\text{fm}^4] $} \\
 &   &$ 0.02  $   &    $ 0.04 $  &  $ 0.08  $ \\
\hline 
\multirow{6}{0.4cm}{\elem{He}{4}}  
& $E_{sum}^{(2)}$ & -19.204 		& -20.269  	&  -23.588  \\
& $E_{sum}^{(3)}$ 	& -20.334 		& -23.224  	&  -26.589  \\
& $E_{sum}^{(10)}$ 	&  -20.507 		& -24.444  	&  -26.947  \\
& $E_{sum}^{(20)}$ 	& -20.526 		& -24.462  	&  -26.964  \\
& $E_{sum}^{(30)}$ 	& -20.537 		& -24.469  	&  -26.971  \\
& CI 	&   -20.539 		& -24.483 	&  -26.994 \\
\hline
\multirow{ 6}{0.4cm}{\elem{O}{16}} 
& $E_{sum}^{(2)}$  & -85.620			& -107.241  	&  -120.699 \\
& $E_{sum}^{(3)}$  & -89.315		& -110.861 	&  -123.863  \\
& $E_{sum}^{(10)}$  & -83.780 		& -107.199 	&  -122.561  \\
& $E_{sum}^{(20)}$	 & -84.180		& -107.341  	&  -122.577  \\
& $E_{sum}^{(30)}$ 	 & -84.018  	& -107.331  	&  -122.577  \\
& CI 				 &  -84.043		& -107.330		&  -122.577 \\
\hline
\multirow{ 6}{0.4cm}{\elem{O}{24}} 
& $E_{sum}^{(2\phantom{})}$  & -125.460		& -124.459  	&  -149.053 \\
& $E_{sum}^{(3)}$  & -122.880 		& -126.670  	&  -151.059  \\
& $E_{sum}^{(10)}$  & -119.705 		& -121.233  	&  -147.446  \\
& $E_{sum}^{(20)}$	 & -119.335 		& -121.314  	&  -147.508  \\
& $E_{sum}^{(30)}$  & -119.483 		& -120.948  	&  -147.489  \\
& CI  	 & -119.131 		& -120.947	&  -147.488 \\
\end{tabular}
\end{ruledtabular}
\end{table}
%%%%%%%%%%%%%%%%%%%%%%%%%%%%%%%%%%%%%%%%%%%%%%%%%%%%%%%%%%%%%%%
%
%
%%%%%%%%%%%%%%%%%%%%%%%%%%%%%%%%%%%%%%%%%%%%%%%%%%%%%%%%%%%%%%%
\begin{figure*}
\centering
\includegraphics[width=\textwidth]{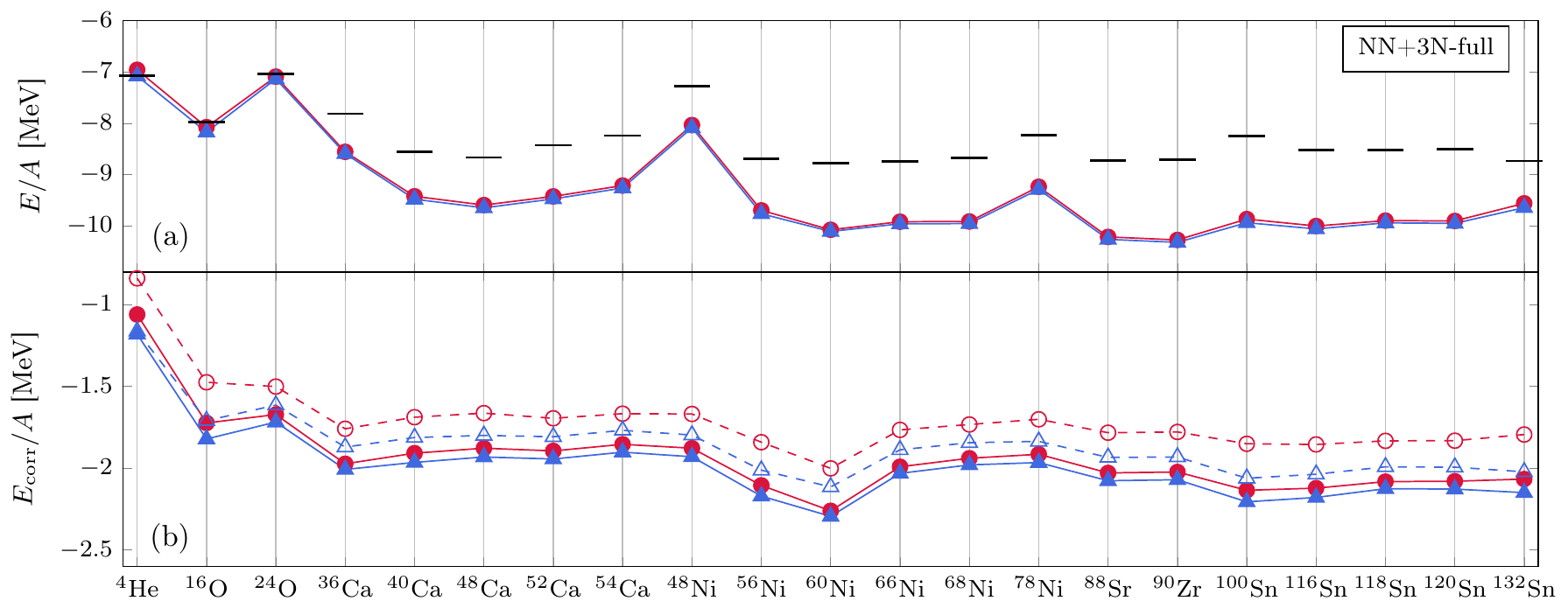}
\caption{Panel (a) shows the ground-state energies per nucleon from third-order HF-MBPT (\redcircle) in comparison to $\text{CR-CC}(2,3)$ (\bluetriangleup) results for selected closed-shell nuclei. Panel (b) shows the correlation energy per nucleon, $E_0^{(2)}$ (\redcircleopen) as well as $E_0^{(2)} + E_0^{(3)}$ (\redcircle) for HF-MBPT. Additionally, the correlation energy per nuclei for CCSD (\bluetriangleupopen) and CR-CC(2,3) (\bluetriangleup) are shown. All calculations were performed with the NN+3N-full interaction with $\alpha=0.08\,\mathrm{\text{fm}^4}$, $\hbar \Omega = 24\,\text{MeV}$ in an $e_{\text{max}}=12$ truncated model space. Experimental values are indicated by black bars.}
\label{V3N}
\end{figure*}
%%%%%%%%%%%%%%%%%%%%%%%%%%%%%%%%%%%%%%%%%%%%%%%%%%%%%%%%%%%%%%%
\begin{figure*}
\centering
\includegraphics[width=\textwidth]{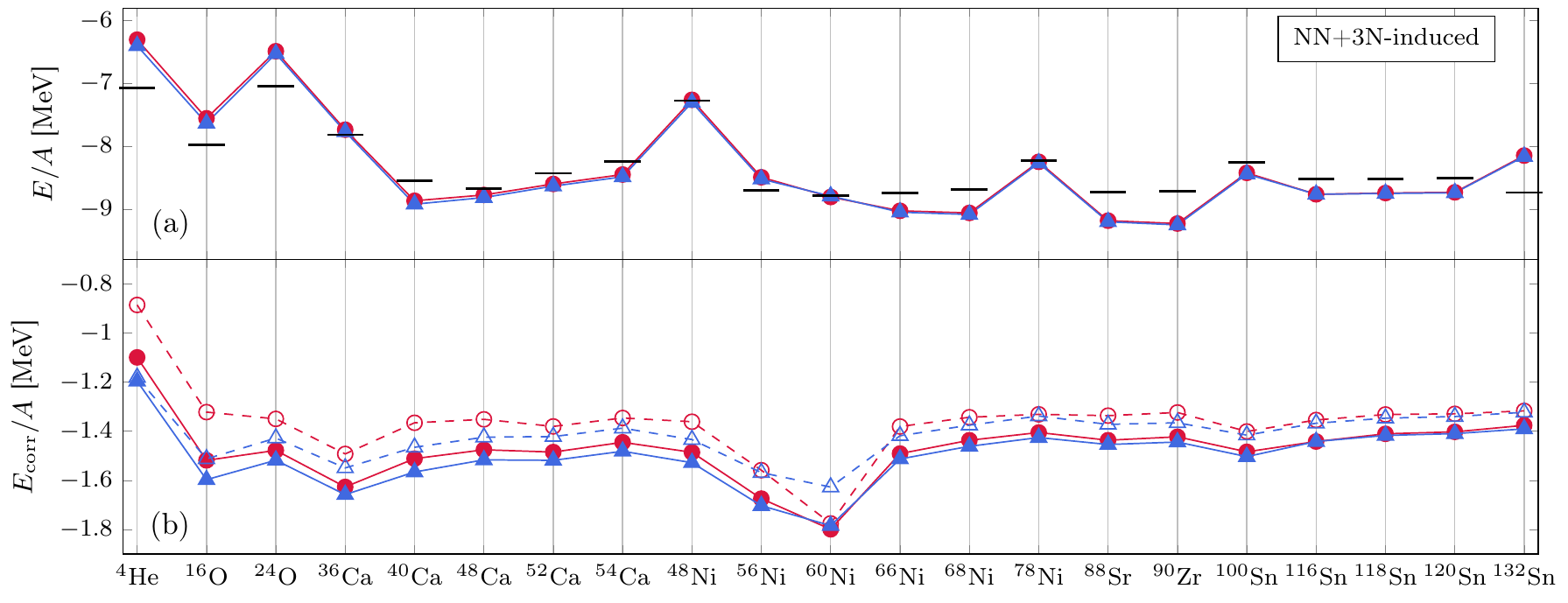}
\caption{Binding energy and correlation energy for the NN+3N-induced interaction. All other parameters as in Fig~\ref{V3N}.
}
\label{VNN}
\end{figure*}
%%%%%%%%%%%%%%%%%%%%%%%%%%%%%%%%%%%%%%%%%%%%%%%%%%%%%%%%%%%%%%%

\paragraph*{Explicit Summation for Heavy Nuclei.}\label{loworder}
For heavier nuclei and larger model spaces we cannot compute the high-order perturbation series explicitly and, thus, we cannot investigate the convergence characteristics explicitly. We can, however, evaluate the perturbative contributions up to third order very efficiently. To demonstrate the validity of a low-order perturbative approximation, we need to compare our results to established \emph{ab initio} techniques, in our case, coupled-cluster calculations with sophisticated triples corrections. 

We consider a sequence of closed-shell nuclei ranging from \elem{He}{4} to \elem{Sn}{132} and perform calculations in second and third-order HF-MBPT in a large model space truncated with respect to the single-particle principal quantum number $e_{\text{max}}=12$. We restrict ourselves to SRG-evolved Hamiltonians with flow parameter $\alpha=0.08\,\text{fm}^4$, which was used extensively in previous calculations and showed favorable order-by-order convergence in our high-order studies. We cannot perform CI calculations for these large spaces, however, the coupled-cluster framework has proven to provide accurate results for ground-state energies of closed-shell nuclei \cite{DeHj04,BaRo07,HaPa10,KoDe04}. We compare the HF-MBPT results to recent CC calculations at the CCSD and the $\text{CR-CC}(2,3)$ level~\cite{PiGo09,BiLa14,Bi14}. Starting from a HF reference state this approach provides a complete inclusion of singly and doubly excited clusters on top of the reference state and, in the case of $\text{CR-CC}(2,3)$ an approximate non-iterative inclusion of triply excited clusters \cite{Ha10,PiWl05,PiWl06,PiWl07}.

In Figs.~\ref{V3N} and \ref{VNN} the ground-state energies per nucleon (a) as well as the correlation energy $E_{\text{corr}} = E-E_{\text{HF}}$ per nucleon (b) from HF-MBPT and $\text{CR-CC}(2,3)$ are depicted for an initial chiral NN+3N and an initial chiral NN interaction. The SRG-induced three-nucleon contribution are taken into account in both cases, leading to the NN+3N-full and NN+3N-induced interactions, respectively. 

These figures show a remarkable result: The binding energies in third-order HF-MBPT and CR-CC(2,3) are in excellent agreement with each other. The relative differences are in most cases much smaller than $1 \%$. The same observation holds for the correlation energy, i.e., the corrections to the HF energy. The third-order energy corrections contribute approximately $0.2$ MeV to the overall binding energy per nucleon and are, therefore, non negligible even though the third-order energy corrections in HF-MBPT are one order of magnitude smaller than the second-order correction. 

%%%%%%%%%%%%%%%%%%%%%%%%%%%%%%%%%%%%%%%%%%%%%%%%%%%%%%%%%%%%%%%
\begin{figure}
\includegraphics[width=0.48\textwidth]{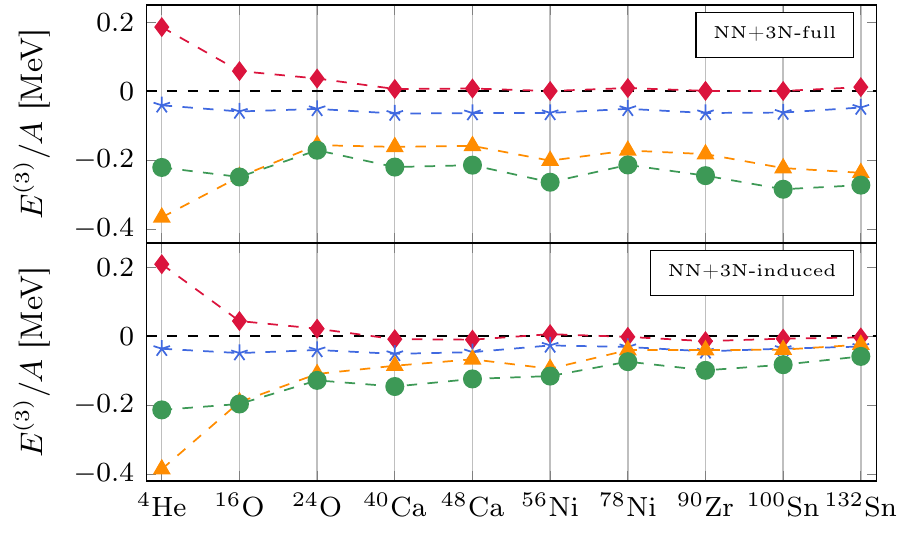}
\caption{Individual contributions of the diagrams appearing at third-order perturbation theory. Show are the contributions per nucleon from the pp-diagram(\reddiamond) , the hh-diagram (\bluestar) and the ph-diagram (\orangetriangleup). The overall contribution of the third-order correction is depicted in (\greencircle).  The first panel corresponds to the NN+3N-full interaction and the second panel to a NN+3N-induced interaction with $\alpha=0.08\,\mathrm{\text{fm}^4}$, $\hbar \Omega = 24\,\text{MeV}$, and $e_{\text{max}}=12$.}
\label{3rdorder}
\end{figure}
%%%%%%%%%%%%%%%%%%%%%%%%%%%%%%%%%%%%%%%%%%%%%%%%%%%%%%%%%%%%%%%

The third-order energy contribution \eqref{eq:thirdorder} consists of three terms corresponding to three Hugenholtz diagrams. Figure~\ref{3rdorder} disentangles their individual contributions to the overall third-order energy correction. The contribution of the $pp$ and $hh$ corrections are almost constant over the entire mass range, whereas the energy correction arising from the $ph$ term scales with increasing mass number in the case of a NN+3N-full interaction. For the tin isotopes the third-order energy correction contributes $3\%$ of the overall binding energy in third-order HF-MBPT and is not negligible. In particular we see that most of the third-order energy correction arises from the $ph$ diagram. In the case of a NN+3N-induced interaction we see that all three terms are suppressed with increasing mass number. These systematic dependencies of the individual third-order contributions on the input Hamiltonian show that a partial inclusion of selected third-order terms may lead to wrong estimates.

%%%%%%%%%%%%%%%%%%%%%%%%%%%%%%%%%%%%%%%%%%%%%%%%%%%%%%%%%%%%%%%

\paragraph*{Conclusions.}\label{conclusion}

We have discussed Rayleigh-Schr\"odinger MBPT as an efficient approach to compute ground-state energies for closed-shell nuclei up to the heavy-mass region. The use of a HF basis has enabled us to overcome convergence problems that generally arise in HO-MBPT. Investigating \elem{O}{16} in different model spaces showed convergent partial sums when using HF-MBPT and the limit of the perturbation series coincides with the results from explicit CI calculations. Additionally, we found systematic dependencies of the convergence rate on the SRG parameter in the case of \elem{O}{16} and \elem{O}{24}.
Thus, in HF-MBPT we can improve the convergence behavior of the perturbation series by further evolving the Hamiltonian, whereas the divergent HO-MBPT series are unaffected. 

We can identify a hierarchy of elements influencing the convergence properties of perturbation series. Defining a partitioning, or equivalently, defining a starting point of the recursive calculation is the most important part.  We have seen from the radically different behavior of the perturbation series in HF-MBPT and HO-MBPT that the partial sums are very sensitive to the partitioning. When using HF-MBPT we can improve the order-by-order convergence by using softer interactions corresponding, e.g., to larger SRG flow parameters. Even for HF basis sets harder interactions can spoil convergence. The `softness' of the interaction has been characterized in terms of Weinberg eigenvalues, which are connected to the spectrum of two-body Green's functions~\cite{W63,Bo06,RaBoFu07}. Similar expressions also appear in the formulas for the first-order state correction. Though the general connection seems obvious, one should be careful with conclusions about the convergence of MBPT for a finite nucleus based on the softness of the interaction. Our work has shown that the partitioning is key for convergence. Our observation that the convergence of HF-MBPT deteriorates for harder interactions could simply be explained by the fact that the unperturbed HF solution becomes a much worse approximation for the ground state in these cases.

The superior convergence properties of HF-MBPT has motivated the use of low-order approximations to investigate nuclei in the medium-mass region. We have validated these low-order approximations to the most sophisticated CC calculations and found excellent agreement of third-order HF-MBPT and CR-CC(2,3) at the level of better than $1\%$. The consistency of high-order partial summations with exact CI diagonalizations as well as the agreement of low-order summations with coupled-cluster results may qualify HF-MBPT as an \emph{ab initio} approach. However, the strong dependence of convergence on the partitioning should be a reason for caution. The HF partitioning seems to be robust for sufficiently soft interactions, but there is no formal guarantee for convergence.  

The great advantage of low-order HF-MBPT is its simplicity: Computationally, the third-order calculations are much cheaper than CC or IM-SRG calculations. It is, therefore, ideal for survey calculations over a large range of medium-mass nuclei, e.g., to explore the ground-state systematics for new interactions.
Formally, the underlying equations and algorithms are trivial compared to CC or IM-SRG.  
As a result of this formal simplicity, extensions to the description exited states and open-shell nuclei are straight-forward. We have demonstrated this already for light nuclei using high-order degenerate HO-MBPT~\cite{LaRo12}. Alternative multi-configurational formulations for open-shell nuclei are under investigation.

\paragraph*{Acknowledgements.}
This work is supported by the DFG through grant SFB 1245, the Helmholtz International Center for FAIR, and the BMBF through contract 05P15RDFN1. S.~Binder gratefully acknowledges the financial support of the Alexander-von-Humboldt Foundation (Feodor-Lynen scholarship).
Numerical calculations have been performed at the computing center of the TU Darmstadt (lichtenberg), at the LOEWE-CSC Frankfurt, and at the National Energy Research Scientific Computing Center supported by the Office of Science of the U.S.~Department of Energy under Contract No. DE-AC02-05CH11231.
%
%
%%%%%%%%%%%%%%%%%%%%%%%%%%%%%%%%%%%%%%%%%%%%%%%%%%%%%%%%%%%%%%%
%

\end{document}